\documentclass[journal]{IEEEtran}
\usepackage[english]{babel}
\usepackage{graphicx}

\newcommand{\ex}[1]{\ensuremath{\times 10^{#1}}}

\begin{document}

\title{Observation and cancellation of the dc Stark shift in strontium optical lattice clocks}
\author{\authorblockN{J\'er\^ome Lodewyck$^1$, Michal Zawada$^{1,2}$, Luca Lorini$^{1,3}$, Mikhail Gurov$^1$ and Pierre Lemonde$^1$}\\
\authorblockA{$^1$LNE-SYRTE, Observatoire de Paris, CNRS, UPMC, 61 avenue de l'Observatoire, 75014 Paris, France}\\
\authorblockA{$^2$Institute of Physics, Nicolaus Copernicus University, Grudziadzka 5, PL-87-100 Torun, Poland}\\
\authorblockA{$^3$Istituto Nazionale di Ricerca Metrologica (INRIM), Strada delle Cacce 91, 10135 Torino, Italy}}

\maketitle

\begin{abstract}
	We report on the observation of a dc Stark frequency shift at the $10^{-13}$ level by comparing two strontium optical lattice clocks. This frequency shift arises from the presence of electric charges trapped on dielectric surfaces placed under vacuum close to the atomic sample. We show that these charges can be eliminated by shining UV light on the dielectric surfaces, and characterize the residual dc Stark frequency shift on the clock transition at the $10^{-18}$ level by applying an external electric field. This study shows that the dc Stark shift can play an important role in the accuracy budget of lattice clocks, and should be duly taken into account.
\end{abstract}

\section{Introduction}

	The development of optical clocks with neutral atoms trapped in an optical lattice has rapidly progressed over the last decade, now surpassing the best microwave cesium clocks with a stability slightly above $10^{-15}/\sqrt{\tau}$ and an accuracy at the $10^{-16}$ level~\cite{KatoPal03, Takamoto05, Ludlow28032008, Baillard08, Lemke09, PTB11}. In addition, a large room for improvement remains, as the stability of lattice clocks is still two orders of magnitude away from the quantum projection noise limit associated with the large number of probed atoms, and the limits in accuracy will likely be pushed in the near future. Optical lattice clocks are therefore good candidates for future optical frequency references. Among them, the development of strontium lattice clocks has received a large attention, with a number of systems already in operation and under construction.

	Many systematic effects in optical lattice clocks arise from the electromagnetic environment of the trapped atoms. The frequency of the atomic transition is displaced by the laser beams forming the optical lattice in which the atoms are confined, through the ac Stark effect, although this displacement is mostly canceled at the `magic' wavelength~\cite{KatoPal03,Takamoto05,PhysRevLett.106.210801}. Other radiations are also involved, such as the clock laser itself, and, more importantly, the blackbody radiation emanating from the vacuum vessel at room temperature. This later systematic effect is now under extensive study~\cite{TacklingBBR} as it is the current limitation to the accuracy of strontium lattice clocks. The influence of the static magnetic field is also considered: the linear Zeeman effect is canceled by alternatively probing magnetic sub-states with opposite quantum numbers $m_F$ split by a bias magnetic field, while its quadratic component is accurately estimated from the measurement of the magnetic field given by the Zeeman splitting. However, the dc Stark shift resulting from the interaction of the atoms with a static electric field had never been considered in the accuracy budget of optical lattice clocks, although an electric field as low as 1~kV/m is enough to induce a shift almost two orders of magnitude larger than the claimed accuracy of optical lattice clocks. Because optical lattice clocks usually require a lot of optical access to cool and trap the atoms, a large solid angle of the environment is made of dielectric material that do not effectively shield the atoms from stray electric fields.

	In this article, we report on the observation, cancellation and characterization of a dc Stark frequency shift detected by comparing the two strontium lattice clocks in operation at LNE-SYRTE. This effect was induced by the presence of electric charges on the surface of curved mirrors placed a few centimeters from the atoms.

This paper is organized as follows. First, we describe the motivations and design goals of optical cavities for the optical lattice in lattice clocks. Then, we report on the observation of a large frequency shift induced by the presence of electric charges on the mirrors of such a cavity. In section~\ref{sec:discharge}, we demonstrate experimental procedures to eliminate and characterize these charges. This papers ends with a prospective study of cavity designs that protect the atoms from electric charges.

\section{Optical lattices formed in optical cavity}

The light shifts induced by the optical lattice is the main specificity of optical lattice clocks. For alkaline-earth atoms considered for optical clocks, the shift associated with the scalar polarizability of the atomic levels involved in the clock transition is by far the largest one, but it is completely canceled at the magic wavelength of the optical lattice. However, other frequency shifts (due to the tensor polarizability, hyper-polarizability and multi-polar polarizabilities~\cite{KatoPal03,Brusch06, PhysRevLett.106.210801}) do not exhibit such a cancellation, and, although they are orders of magnitude smaller than the scalar shift, require a well-engineered optical lattice in terms of power and polarization in order to characterize and control them. For this purpose, lattices in optical cavities provide for a clean and reproducible trapping potential for the atoms. Furthermore, the power build-up in the resonator enables us to reach very large trap depths thanks to which we could recently measure the various lattice light shifts for strontium atoms to an unprecedented level~\cite{PhysRevLett.106.210801}. This study tackled many of the main effects that used to limit the accuracy of strontium lattice clocks. We expect that similar optical cavities will be implemented in other lattice clocks to measure the light shift effects with other atomic species.

We have built two optical lattice cavities at LNE-SYRTE, with different cavity designs. The first one consists of a L shaped resonator with two meniscus mirrors of curvature radius 200 and 125 mm separated by 320 mm. Inside the cavity lay the two windows of the vacuum chamber and a 45 plate playing the role of a polarization discriminator. The eigen-mode of the cavity has a waist of 89~$\mu$m where the Sr atoms are trapped. The finesse of the cavity is 80 and the intra-cavity optical power goes up to 12 W. The resonance frequencies of the two polarization eigen-modes of the cavity are distinct enough to ensure a stable polarization with a pure linearity ($ > 99.9$\%  in terms of electric field). The length of the cavity is locked with a H\"ansch Couillaud (HC) system to keep the laser light on resonance.

For the second clock a rather different design has been chosen. We opted for a linear cavity entirely placed under vacuum with a length of 60~mm. The plane-concave mirrors are used as the vacuum chamber windows and are articulated through flexible bellows. Their curvature radius is 35 mm, thus forming a lattice with a 56~$\mu$m waist and a finesse of 150. The length of the cavity is locked by a Pound-Drever-Hall (PDH) scheme. This newer design brings a number of improvements over the first design:
\begin{itemize}
	\item The line-width of the cavity is larger, thus dramatically reducing the efficiency of the laser frequency-to-amplitude noise conversion by the cavity. Indeed, the amplitude noise in the cavity parametrically heats the atoms and eventually expels them from the trapping potential~\cite{PhysRevA.56.R1095}. With the first cavity design, a pre-stabilization of the tapping laser frequency on a reference cavity is required, and the trapping lifetime is limited to about 500~ms. With the second design,  the parametric heating is no more a limit for the lifetime of the trapped atoms.
	\item The polarization eigen-modes are degenerated, which allows the complete study of the polarization dependence of the polarizability and hyper-polarizability of the clock transition.
	\item Because of the absence of intra-cavity optical elements, we have a unity impedance matching of the input beam with the cavity. Combined with a higher finesse and a smaller waist, higher trap depths can be reached.
	\item A short cavity placed under vacuum enables a solid design insensitive to mechanical drifts and dust accumulation. Moreover, the PDH lock requires less tuning than the HC scheme. With this configuration, the long term operation of the clock is more reliable and suits more efficiently the long integration times usually required for metrology experiments. Effectively, this cavity has been in operation for many months with only minor realignment.
\end{itemize}

Optical lattices in cavities enables us to access very large trapping depths. With the first clock, depths as large as $U = 1000$~$E_R$ ($E_R$ being the recoil energy associated with the absorption of one lattice photon) could be achieved. In the second clock, with its finer cavity, trapping potentials as deep as 2600~$E_R$ were observed. These depths are ideally suited to explore the polarizability and hyper-polarizability of the clock transition. Moreover, with such large trap depths, the second stage Magento-Optical Trap (MOT) usually used to bridge the temperature gap between the Sr MOT using the intense $^1S_0 \longrightarrow {^1}P_1$ transition at 2~mK and the shallow dipole-trapping depth is not required. Instead, we use the `drain' technique~\cite{Brusch06} in which the deep lattice is overlapped with the optical lattice as well as two laser beams tuned on the $^1S_0 \longrightarrow {^3}P_1$ (689~nm) and $^3P_1 \longrightarrow {^3}S_1$ (688~nm) transition with a waist slightly smaller than the optical lattice. With these two beams, the atoms from the MOT that cross the lattice position are drained in the ${^3}P_0$ and ${^3}P_2$ metastable state and therefore escape the MOT dynamics. Among these, the coldest fraction accumulate in the lattice dipole trap. Within a few 100~ms, about $10^4$ atoms are loaded in the optical lattice, scaling as $U^{3/2}$.

\section{Observation of electric charges on the cavity mirrors}

\begin{figure}
	\begin{center}
		\includegraphics[width=0.5\textwidth]{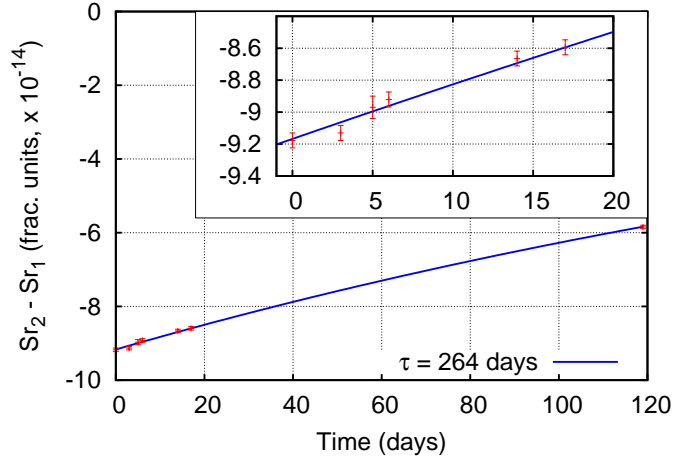}
	\end{center}
	\caption{\label{fig:discharge}Temporal decrease of the frequency difference between the two strontium clocks. The exponential decay is due to the slow discharge of the charges through the mirror substrate. We measured a time constant of 264 days for the frequency decay corresponding to 528 days for the charge because of the quadratic nature of the Stark effect. Knowing the static polarizability $\alpha$ of the strontium clock transition~\cite{PhysRevA.78.032508}, we can deduce the magnitude of the electric field at level of the atoms of 3.4~kV/m at $t$ = 0. Inset: closer view on the first points. On a short time scale, no variation of the frequency shift was observed as the frequency difference between the two clocks averages down with a statistical uncertainty below $10^{-16}$ after one hour of measurement.}
\end{figure}

Upon completion of the in-vacuum cavity for the second clock, a frequency difference on the order of $10^{-13}$ was observed between the two strontium lattice clocks, orders of magnitude higher than the realized and expected accuracies of lattice clocks. This frequency difference has been attributed to the dc Stark shift induced by the presence of electric charges on the reflecting surface of the cavity mirrors. Quantitatively, if $\delta E$ is the norm of the electric field induced by the charges, the stark shift $\delta \nu$ on the clock frequency reads:
\begin{equation}
	\delta \nu = -\frac 1 2 \alpha (\delta E)^2,
\end{equation}
where $\alpha$ is the difference between the static polarizabilities of the excited and fundamental states of the clock transition.

Over time, these charges slowly migrate through the mirror substrate to eventually reach the opposite plane surface itself connected to the electrical ground. Figure~\ref{fig:discharge} shows the effect of the exponential discharge of the mirrors with a time constant of 528 days. On the short term, however, the stark shift shows itself very stable as it does not degrade the stability of the frequency difference between the two clocks over time scales of a few $10^4$ seconds. In addition, it shows that building two similar frequency references is a crucial test for establishing the accuracy budget of the clocks.

By finite element modeling (FEM) of the mirrors geometry and ground planes of the vacuum chamber, we estimate the electric capacity of the optical cavity on the order of $1$~pF, the difference of electric potential between the mirrors on the order of $1$~kV, which yields an electric charge of $\pm 1 $~nC. Subsequently, from the time constant of the discharge, we can estimate the resistivity of the mirror substrate at the level of $10^{18}$~$\Omega$m. The resistivity of glass is mainly determined by the amount of impurities it embeds, and therefore can span several orders of magnitude, from $10^{15}$~$\Omega$m to $10^{18}$~$\Omega$m. However, optical grade glasses usually feature the least impurity hence the highest resistivity, in agreement with our model and measurements.

The origin of the charges trapped on the mirrors is unclear, as we could not purposely charge them in a reproducible manner. These charges must originate from static electricity accumulated while the mirrors were assembled, because we observed that the pumping process of the vacuum chamber from atmospheric pressure does not recharge the mirrors.

Systematic shifts induced by patch charges is a known effect for ion clocks, in which the isolating dielectric of the electrodes or the collecting optics can easily be charged by the ionizing process. Fortunately, the motion of the ion usually reveals the presence of these charges~\cite{ions}. To the best of our knowledge, this papers reports the first observation of the effect of electric charges in neutral optical clocks. Furthermore, the large magnitude of the effect, four orders of magnitude higher than the targeted accuracy of strontium clocks raises some concerns whether this effect might reveal itself to be a show-stopper for the further development of lattice clocks. Indeed, although an optical cavity close to the atoms may be an extreme scenario, the operation of optical lattice clocks requires a large optical access where dielectric component are prone to charge accumulation. In the following sections, we describe experiments we have conducted to eliminate and characterize the residual amount of charges as well as prospects for further reduction, if required.

\section{Discharge and measurement of patch charges}
\label{sec:discharge}

\begin{figure}
	\begin{center}
		\includegraphics[width=0.5\textwidth]{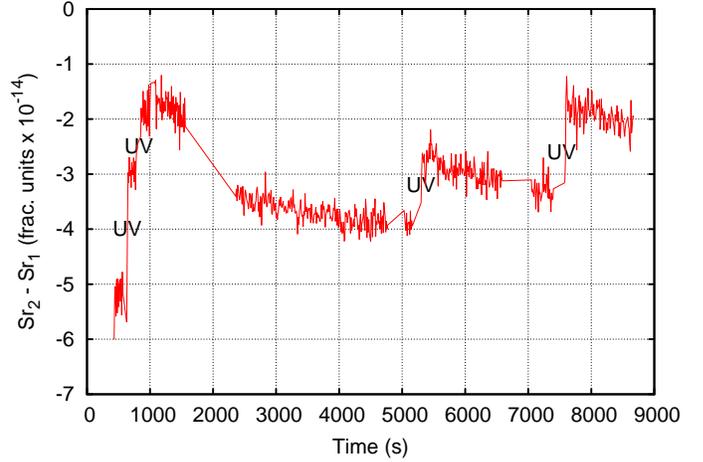}
	\end{center}
	\caption{\label{fig:UV}Frequency difference between the two strontium lattice clocks. Each `UV' tag represents a 30~s UV illumination with a broadband UV lamp centered on 360~nm distributed over a large solid angle. After each illumination, the Stark shift is halved, but a fast relaxation of the charges is observed shortly after (note that at 2000~s, the lattice wavelength was stabilized on the magic wavelength introducing a gap of several Hz in the frequency difference). After the illumination with a UV LED centered on 365~nm and a power of about 100~mW during 30 hours, the two strontium clock frequency matched at the $10^{-16}$ level.}
\end{figure}

The removal of trapped charges from the insulated surface of the mirrors inside the vacuum chamber was achieved by shining UV light on the mirrors through the vacuum chamber optical accesses. Such a technique had been successfully developed and characterized for gravitational wave detectors~\cite{PhysRevD.81.021101}. The most probable mechanism is the extraction of charges from the surface by the photoelectric effect. The charges would then follow the electric field lines to the ground. Another mechanism based on the creation of unstable paramagnetic centers in silica has also been reported~\cite{Ugolini20085741}, but given the thickness of the mirrors and the discharge rate we obtained, this effect is unlikely to occur in our case. Figure~\ref{fig:UV} shows the reduction of the Stark shift on the clock transition after 2 minutes of illumination with UV light down to the $10^{-15}$ level. After keeping the UV light for two consecutive nights, the frequency difference between the clocks became compatible with zero within the accuracy budget of the strontium lattice clocks on the order of $10^{-16}$.

\begin{figure}
	\begin{center}
		\includegraphics[width=0.5\textwidth]{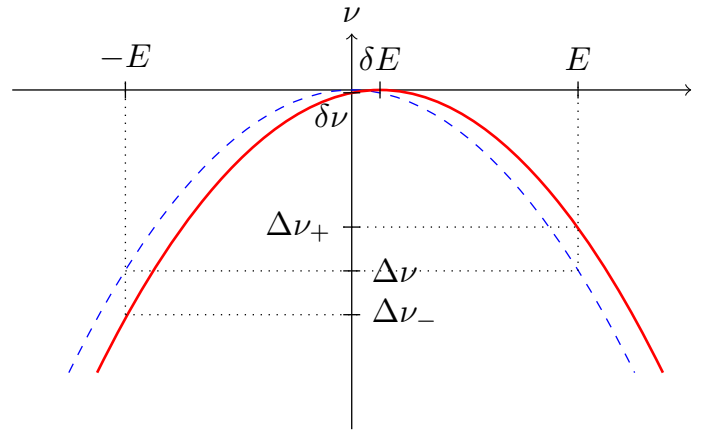}
	\end{center}
	\caption{\label{fig:stark}DC Stark shift of the frequency of the clock transition as a function of the electric field generated by an electrode placed in the vicinity of the atoms. The dashed parabola shows the Stark shift on the clock frequency without any residual field. A residual field $\delta E$ due to charges slightly displaces the center of the parabola (thick solid parabola), thus introducing a difference $\Delta \nu_+ - \Delta \nu_-$ in the frequency of the clock transition when probed with opposite voltages applied on the electrode. This sketch illustrates the large leverage of this technique: even a minute frequency shift $\delta\nu$ on the clock transition in the absence of electrode voltage would translate in a large frequency unbalance in the presence of the electric field $E$.}
\end{figure}

In spite of the apparently fast discharge of the cavity mirrors and the good agreement of the two strontium clocks, a measurement of the electric field at the level of the atoms is still required to independently estimate the residual magnitude of the dc Stark effect that should appear in the clock accuracy budget. For this, we creates an electric field $E$ at the level of the atoms with an electrode placed a few centimeters away from the atoms, outside the vacuum chamber, which . A voltage of $V = 1$~kV on the electrode is enough to induce a shift $\Delta \nu = 26$~Hz (or $\Delta \nu/\nu = 6\ex{-14}$ in fractional units, $\nu = 429.228$~THz being the Sr clock frequency) on the frequency of the clock transition. Because of the quadratic behavior of the Stark effect, this frequency shift should stay constant when the voltage of the electrode is reversed. If not, it reveals a residual electric field $\delta E$ in the direction of the applied field $E$~\cite{Matveev:11} (see figure~\ref{fig:stark}). Quantitatively, the residual systematic frequency shift $\delta \nu$ due to the residual field $\delta E$ reads:
\begin{equation}
	\delta \nu = \frac{\left(\Delta \nu_+ - \Delta \nu_-\right)^2}{16\Delta \nu},
\end{equation}
where $\Delta \nu_{\pm}$ is the Stark shift on the clock frequency when the voltage $\pm V$ is applied on the electrode. This expression is remarkable in the sense that only the difference between the Stark shifts needs to be measured with a good precision. Furthermore we can see that the average Stark shift $\Delta \nu$ acts as a leverage to measure $\delta \nu$ with a good accuracy without requiring large integration times. Finally, when switching the polarity of the electrode, the difference $\delta V$ between the absolute values of the voltage has to fulfill $\delta V/V \ll \delta E/E$ so that the voltage imbalance $\delta V$ is not interpreted as a residual electric field. In our experiment, we have $\delta V/V \simeq 10^{-5}$, which is sufficient for all practical purposes.

For an electric field $E$ non-orthogonal to the axis of the cavity, we measured $\Delta \nu_+ - \Delta\nu_- = 0.1 \pm 0.4$~Hz (or $(\Delta \nu_+ - \Delta\nu_-)/\nu=(2\pm9)\ex{-16}$ in fractional units) after an integration time of 400~s. Taking into account the angle between the electrode and the cavity, this yields a residual systematic frequency shift due to the charges compatible with zero with an uncertainty well within the target accuracy for optical lattice clocks:
\begin{equation}
	\frac{\delta \nu}{\nu} = 0 \pm 1.5\ex{-18}.
\end{equation}
To be complete, this experiment should be reproducted with three non-orthogonal directions of the applied electric field $E$. However, because the electrode has been carefully placed so that the electric field is not orthogonal to any specific plane on the vacuum vessel, and because the UV light is applied in an isotropic way, we can safetly conclude that this experiment shows that the UV light very efficiently removed the trapped charges.

Interestingly enough, after 400~s of integration, we could observe a revival of charges with a random walk behavior. After one hour, this new charge accumulation induced an electric field of about $130$~V/m at the level of the atoms giving a Stark shift of $\delta \nu/\nu = 1.5\ex{-16}$. This accumulation can be dramatically enhanced by the same UV light used for discharging when shone in conjunction with the electric field $E$. This observation backs the photo-electric interpretation of the effect of the UV light: UV photon can extract charges from the vacuum chamber material, and these charges follow the electric field lines, in this later case generated by the electrode. However, these charges rapidly disappeared within a few minutes after the electric field $E$ and UV light were switched off, indicating that they were accumulating on a dielectric surface with a larger conductivity than the cavity mirrors, most probably the vacuum chamber window close to the electrode.

\section{Shielding the atoms from stray electric fields}

\begin{figure}
	\begin{center}
		\includegraphics[width=0.5\textwidth]{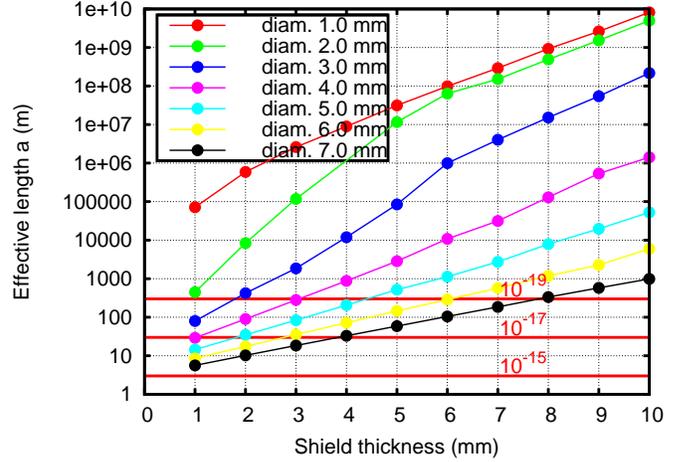}
	\end{center}
	\caption{\label{fig:shield}Effective electrical length of the optical lattice cavity for different shielding geometries. A longer length means a small electric field experienced by the atoms. As expected, the effective length rises when the shield thickness increases or when the shield aperture reduces. The graph indicates the accuracy levels reached for given effective lengths assuming that the cavity mirrors are charge as we initially observed in our experiment. We can observe that the charges are efficiently screened, even for casual geometries of the shielding cylinder.}
\end{figure}

Although we have seen in the previous section that charges could be easily removed using UV light, a better design of the vacuum chamber in which the atoms are shielded from the dielectric surfaces of the cavity mirrors could help to avoid the burden of repeatedly checking that no dc Stark shifts impairs the accuracy of the clock. For this purpose, we modified our finite element modeling of the vacuum chamber with the addition of grounded cylindrical pieces, made of conductive metal, that cover the surface of the cavity mirrors. These pieces are characterized by their thickness and by the inner diameter of the cylinder through which the lattice light reaches the mirrors. The FEM simulation gives the effective length of the cavity defined as the ratio between the difference of electric potential on the surface of the mirrors (imposed by the amount of charges trapped on the mirrors) and the electric field experienced by the atoms. For the unshielded cavity, the effective length is 30~cm. Figure~\ref{fig:shield} shows that a conductive cylinder with a 4~mm internal diameter and a thickness of 4~mm would be sufficient to reduce the Stark shift from $10^{-13}$ to $10^{-19}$.

A more drastic method to get rid of charge accumulation on dielectric materials would be to cover the entire surface of the vacuum chamber with a thin layer of titanium using a sublimation pump. This thin layer would make the surface conductive, efficiently driving the charges to the ground. However, it is unclear whether such a process would preserve the finesse of the optical cavity.

\section{Conclusion}

In this paper, we have proposed the first study of the dc Stark effect in optical lattice clocks. We have demonstrated that charges can accumulate to the point where they induce a shift orders of magnitude larger that the claimed clock accuracy. Then, we have proposed efficient methods to remove and characterize these charges. This study shows that the Stark effect, when carefully controlled, can play a negligible role in the clock accuracy budget, at the level of $10^{-18}$.

This work has received funding from the European Community's FP7, ERA-NET Plus, under Grant Agreement No. 217257, as well as from CNES and ESA (SOC project)



\begin{thebibliography}{10}
\providecommand{\url}[1]{#1}
\csname url@samestyle\endcsname
\providecommand{\newblock}{\relax}
\providecommand{\bibinfo}[2]{#2}
\providecommand{\BIBentrySTDinterwordspacing}{\spaceskip=0pt\relax}
\providecommand{\BIBentryALTinterwordstretchfactor}{4}
\providecommand{\BIBentryALTinterwordspacing}{\spaceskip=\fontdimen2\font plus
\BIBentryALTinterwordstretchfactor\fontdimen3\font minus
  \fontdimen4\font\relax}
\providecommand{\BIBforeignlanguage}[2]{{%
\expandafter\ifx\csname l@#1\endcsname\relax
\typeout{** WARNING: IEEEtran.bst: No hyphenation pattern has been}%
\typeout{** loaded for the language `#1'. Using the pattern for}%
\typeout{** the default language instead.}%
\else
\language=\csname l@#1\endcsname
\fi
#2}}
\providecommand{\BIBdecl}{\relax}
\BIBdecl

\bibitem{KatoPal03}
H.~Katori, M.~Takamoto, V.~G. Pal'chikov, and V.~D. Ovsiannikov, ``Ultrastable
  optical clock with neutral atoms in an engineered light shift trap,''
  \emph{Phys. Rev. Lett.}, vol.~91, p. 173005, 2003.

\bibitem{Takamoto05}
M.~Takamoto, F.-L. Hong, R.~Higashi, and H.~Katori, ``An optical lattice
  clock,'' \emph{Nature}, vol. 435, p. 321, 2005.

\bibitem{Ludlow28032008}
A.~D. Ludlow, T.~Zelevinsky, G.~K. Campbell, S.~Blatt, M.~M. Boyd, M.~H.~G.
  de~Miranda, M.~J. Martin, J.~W. Thomsen, S.~M. Foreman, J.~Ye, T.~M. Fortier,
  J.~E. Stalnaker, S.~A. Diddams, Y.~Le~Coq, Z.~W. Barber, N.~Poli, N.~D.
  Lemke, K.~M. Beck, and C.~W. Oates, ``Sr lattice clock at $1\times 10^{-16}$
  fractional uncertainty by remote optical evaluation with a ca clock,''
  \emph{Science}, vol. 319, no. 5871, pp. 1805--1808, 2008.

\bibitem{Baillard08}
X.~Baillard, M.~Fouch{\'e}, R.~{Le Targat}, P.~Westergaard, A.~Lecallier,
  F.~Chapelet, M.~Abgrall, G.~Rovera, P.~Laurent, P.~Rosenbusch, S.~Bize,
  G.~Santarelli, A.~Clairon, P.~Lemonde, G.~Grosche, B.~Lipphardt, and
  H.~Schnatz, ``An optical lattice clock with spin-polarized {$^{87}$Sr}
  atoms,'' \emph{Eur. Phys. J. D}, vol.~48, pp. 11--17, 2008.

\bibitem{Lemke09}
N.~D. Lemke, A.~D. Ludlow, Z.~W. Barber, T.~M. Fortier, S.~A. Diddams,
  Y.~Jiang, S.~R. Jefferts, T.~P. Heavner, T.~E. Parker, and C.~W. Oates,
  ``Spin-1/2 optical lattice clock,'' \emph{Phys. Rev. Lett.}, vol. 103, no.~6,
  p. 063001, 2009.

\bibitem{PTB11}
J.~V. W. T. M. S. V. S. W. B. L. G. G. F. R. U.~S. St.~Falke, H.~Schnatz and
  C.~Lisdat, ``The {$^{87}$Sr} optical frequency standard at {PTB},''
  \emph{arXiv:1104.4850}, 2011.

\bibitem{PhysRevLett.106.210801}
P.~G. Westergaard, J.~Lodewyck, L.~Lorini, A.~Lecallier, E.~A. Burt, M.~Zawada,
  J.~Millo, and P.~Lemonde, ``Lattice-induced frequency shifts in sr optical
  lattice clocks at the $10^{-17}$ level,'' \emph{Phys. Rev. Lett.}, vol. 106,
  no.~21, p. 210801, May 2011.

\bibitem{TacklingBBR}
S.~F. J. V. W. F.~R. Th.~Middelmann, Ch.~Lisdat and U.~Sterr, ``Tackling the
  blackbody shift in a strontium optical lattice clock,'' \emph{IEEE Trans. on
  Instrum. and Meas.}, vol.~60, p. 2550, 2011.

\bibitem{Brusch06}
A.~Brusch, R.~{Le Targat}, X.~Baillard, M.~Fouch{\'e}, and P.~Lemonde,
  ``Hyperpolarizability effects in a {Sr} optical lattice clock,'' \emph{Phys.
  Rev. Lett.}, vol.~96, p. 103003, 2006.

\bibitem{PhysRevA.56.R1095}
T.~A. Savard, K.~M. O'Hara, and J.~E. Thomas, ``Laser-noise-induced heating in
  far-off resonance optical traps,'' \emph{Phys. Rev. A}, vol.~56, no.~2, pp.
  R1095--R1098, Aug 1997.

\bibitem{PhysRevA.78.032508}
S.~G. Porsev, A.~D. Ludlow, M.~M. Boyd, and J.~Ye, ``Determination of sr
  properties for a high-accuracy optical clock,'' \emph{Phys. Rev. A}, vol.~78,
  no.~3, p. 032508, Sep 2008.

\bibitem{ions}
W.~M.~I. M.~E.~Poitzsch, J. C.~Bergquist and D.~J. Wineland, ``Cryogenic linear
  ion trap for accurate spectroscopy,'' \emph{Rev. of Sci. Instrum.}, vol.~67,
  p. 129, 1996.

\bibitem{PhysRevD.81.021101}
S.~E. Pollack, M.~D. Turner, S.~Schlamminger, C.~A. Hagedorn, and J.~H.
  Gundlach, ``Charge management for gravitational-wave observatories using uv
  leds,'' \emph{Phys. Rev. D}, vol.~81, no.~2, p. 021101, 2010.

\bibitem{Ugolini20085741}
D.~Ugolini, M.~Girard, G.~Harry, and V.~Mitrofanov, ``Discharging fused silica
  test masses with ultraviolet light,'' \emph{Phys. Lett. A}, vol. 372, no.~36,
  pp. 5741 -- 5744, 2008.

\bibitem{Matveev:11}
A.~Matveev, C.~G. Parthey, A.~Beyer, N.~Kolachevsky, J.~Alnis, R.~Pohl,
  T.~Udem, and T.~W. Haensch, ``Systematic frequency shifts in spectroscopy of
  1s-2s transition in atomic hydrogen,'' 2011.

\end{thebibliography}
\end{document}